\documentclass[11pt]{article}
\usepackage[utf8]{inputenc} 
\usepackage[english]{babel} 
\usepackage{geometry} 
\usepackage{graphicx} 
\usepackage{float} 
\usepackage{longtable} 
\usepackage{listings} 
\usepackage{xcolor} 
\usepackage{enumitem} 
\usepackage{multirow} 
\usepackage{pdflscape} 
\usepackage{moresize} 
\usepackage{amsmath, amsthm, amssymb} 
\usepackage{algorithm} 
\usepackage{algpseudocode} 
\usepackage{array} 
\usepackage{tabularx}
\usepackage{changepage} 
\usepackage{booktabs} 
\usepackage{pifont}
\usepackage{url} 
\usepackage{diagbox}
\usepackage[colorlinks=true, citecolor=blue, linkcolor=blue, urlcolor=blue]{hyperref} 
\usepackage{makecell}
\usepackage[table]{xcolor}
\usepackage{booktabs, multirow}

\definecolor{highlight}{RGB}{204,229,255} 

\title{\textbf{Empirical parameterization of the Elo Rating System}}
\author{
\small
\begin{tabular}{@{}c@{\hspace{0.7cm}}c@{\hspace{0.7cm}}c@{}}
\textbf{Shirsa Maitra} & \textbf{Tathagata Banerjee} & \textbf{Anushka De} \\
\textit{Computer Science Engineering} & \textit{Dept. of Statistics \& Data Science} & \textit{Dept. of Statistics \& Data Science} \\
\textit{Heritage Institute of Technology, Kolkata} & \textit{National University of Singapore} & \textit{Northwestern University, Illinois} \\
\end{tabular}
\\
\\
\small
\begin{tabular}{@{}c@{\hspace{1cm}}c@{}}
\textbf{Diganta Mukherjee} & \textbf{Tridib Mukherjee} \\
\textit{Sampling \& Official Statistics Unit (SOSU)} & \textit{Chief Data Scientist \& AI Officer} \\
\textit{Indian Statistical Institute, Kolkata} & \textit{IDfy, India} \\
\end{tabular}
}

\date{}
\begin{document}
\maketitle
\begin{abstract}
    This study aims to provide a data-driven approach for empirically tuning and validating rating systems, focusing on the Elo system. Well-known rating frameworks, such as Elo, Glicko, TrueSkill systems, rely on parameters that are usually chosen based on probabilistic assumptions or conventions, and do not utilize game-specific data. To address this issue, we propose a methodology that learns optimal parameter values by maximizing the predictive accuracy of match outcomes. The proposed parameter-tuning framework is a generalizable method that can be extended to any rating system, even for multiplayer setups, through suitable modification of the parameter space. Implementation of the rating system on real and simulated gameplay data demonstrates the suitability of the data-driven rating system in modeling player performance. 
\end{abstract}


\section{Introduction}
    A key component of sports analytics is determining a player's skill level in a game. Fair matchmaking, ranking and performance tracking of players and competitive integrity all depend on the quantification of player skill. Accurate skill estimation is also crucial for fraud detection, quantifying whether a game is skill or luck-based~\cite{banerjee2025skillvschance}, to improve player engagement, and creating balanced gaming environments in professional sports and online platforms. Many methods have been developed over time to measure player skill, and the majority of these methods rely on performance-based metrics derived from gameplay outcomes. The Elo rating system~\cite{arpad_elo} has gained widespread acceptance as a standard for allocating numerical ratings that represent a player's relative skill in two-player games like chess. After every game, these ratings are updated to reflect a player's dynamic skill level more accurately. 
    
    The Glicko~\cite{Glicko_original} and Glicko-2~\cite{glicko2example} systems are two examples of extensions of Elo that have been created and are in use in practice. These systems use rating volatility and confidence intervals to more accurately capture uncertainty over longer time periods.
    
    In this study, we propose a data-driven approach to parameter tuning in the Elo rating system. Glickman (1999)~\cite{glicko_parameter_tuning} introduced a state-space approach for dynamic parameter estimation, treating player strength as a time-evolving latent variable. While his work was focused on probabilistic estimation of skill volatility and uncertainty, our method takes a complementary, generalized empirical approach of optimized parameter tuning directly from data based on the predictive power~\cite{football_pred, kovalchik2020elo_mov} of player ratings. This framework can  be extended to any rating system with finite parameters, such as Glicko or TrueSkill(and its advancements and variations)~\cite{trueskill_bayesian, minka2018trueskill2, trueskill_through_time}, where only the underlying formulas and the parameter space change, but the optimization strategy remains the same. The  resultant rating system is then implemented in simulated and real gameplay datasets to study its applicability in player performance modeling. 
    
    This paper puts forward a comprehensive study describing the methodology of the Elo rating system for two-player games and suggests a detailed framework for parameter tuning in rating system formulations. This enables enhancing the sensitivity of rating models to specific game environments. Section~\ref{the_elo_rating_system} begins with a brief background on the Elo rating system, followed by a short subsection discussing the key assumptions underlying different rating models. This is followed by a detailed formulation of the Elo update mechanism in a two-player setting and introduces a sensitivity factor based on player experience to improve rating stabilization. Section~\ref{our_contribution} outlines the core contribution, which is our proposed parameter tuning approach, supported by concrete reasoning for our assumptions, choice of classification models, and parameters. The empirical analysis is conducted using the proposed framework on both real and simulated gameplay datasets in Section~\ref{empirical_results}. This section also serves as an illustrative example of how parameter configurations can be evaluated and their values can be optimized using our parameterization technique to identify the best values for a given rating model. Section~\ref{illustration_of_results} discusses the more in-depth findings and interpretations of the empirical study. We conclude the paper by providing a summary of the study's objectives and outlining possible avenues for further research that could build on this work in section \ref{conclusion}.

\section{The Elo Rating System}
\label{the_elo_rating_system}
\subsection{Background}

The Elo rating system was developed by Arpad Elo, a Hungarian-American physicist and chess master, as a mathematical method to measure the relative skill levels of players in zero-sum games. Introduced by the United States Chess Federation in 1960 and later adopted by FIDE, it replaced older, subjective ranking systems with a probabilistic model based on expected outcomes of matches. The system assumes that player performance follows a normal distribution and updates ratings dynamically based on match results. Its simplicity and accuracy made it the foundation of modern competitive ranking systems used in chess, e-sports, and various online platforms.
\subsection{Assumptions in Rating System Models} 
Most rating systems rely on specific assumptions that influence how player strength is realized from match outcomes.

\begin{itemize}
    \item Elo Rating System: The Elo system~\cite{arpad_elo} assumes that player performance follows a logistic distribution centered around their latent skill level. If $R_A$ and $R_B$ denote the ratings of players $A$ and $B$, the expected score for $A$ is given by,
    
    $$E_A = \frac{1}{1 + 10^{(R_B - R_A)/400}}$$

    After each game, the updated rating is computed as,
    $$R_A' = R_A + K (S_A - E_A)$$
    where $S_A \in \{0, 0.5, 1\}$ represents the actual outcome and $K$ is a sensitivity factor. Elo assumes static skill between games and does not explicitly model uncertainty or volatility.
    
    \item Glicko Models: The Glicko models~\cite{Glicko_original, glicko2example} extends Elo by introducing a rating deviation $(RD)$, representing uncertainty and variability in a player’s skill estimate. Player skill is assumed to follow a Gaussian distribution,
    $\theta_i \sim \mathcal{N}(\mu_i, \sigma_i^2)$,
    where $\mu_i$ is the mean rating and $\sigma_i$ evolves with inactivity and game outcomes. Glicko-2 further introduces a volatility parameter $\tau$ that governs how rapidly skill can change over time, adapting to performance variance.
    
    \item Microsoft's TrueSkill: TrueSkill~\cite{trueskill_bayesian} generalizes the Elo and Glicko systems through a Bayesian framework. Instead of assigning a fixed rating, TrueSkill adopts a Bayesian approach, modelling each player's skill as a Gaussian random variable,
    $p_i \sim \mathcal{N}(\mu_i, \sigma_i^2)$,
    where $\mu_i$ represents the estimated skill mean and $\sigma_i$ encodes uncertainty about that estimate.  
    In a two-player setting, the probability that player $i$ defeats player $j$ is given by:
    $$P(i \text{ wins}) = \Phi \!\left( \frac{\mu_i - \mu_j}{\sqrt{2(\sigma_i^2 + \beta^2)}} \right)$$
    where $\Phi(\cdot)$ is the cumulative density of the standard normal distribution, and $\beta$ represents the random performance variance during a game.  
    After each match, both $\mu_i$ and $\sigma_i$ are updated using approximate Bayesian updation, allowing the model to dynamically capture improvement, learning, and uncertainty reduction over time. Unlike Elo or Glicko, which assume scalar and independent rating updates, TrueSkill naturally extends to multiplayer or team-based games by inferring posterior skill distributions jointly from all participants’ outcomes. TrueSkill-2~\cite{minka2018trueskill2} further refines this process by enabling multidimensional skill vectors and real-time volatility adaptation, at the cost of increased computational complexity and reduced interpretability.
\end{itemize}

\subsection{Formulation}

The symbols and parameters used in this section and the following sections are summarized below.

\begin{table}[H]
\centering
\renewcommand{\arraystretch}{1.2}
\begin{tabular}{ll}
\hline
\textbf{Symbol} & \textbf{Description} \\
\hline
$R_0$ & Initial rating assigned to a player upon registration \\
$rating_1, rating_2$ & Current ratings of Player 1 and Player 2 before the match \\
$rating_1', rating_2'$ & Updated ratings of Player 1 and Player 2 after the match \\
$K_1, K_2$ & Sensitivity or learning rate constants for Players 1 and 2 \\
$S_1, S_2$ & Actual scores of Players 1 and 2 ($1 =$ win, $0.5 =$ draw, $0 =$ loss) \\
$E_1, E_2$ & Expected scores of Players 1 and 2, computed from the rating difference \\
$D$ & Scaling factor controlling the shape of the logistic function (commonly 400) \\
$n_i$ & Number of games played by player $i$, used to adapt $K_i$ \\
$K_a, K_b, K_c$ & Rating update constants for early, mid, and late stages of player experience \\
$n_{c_1}, n_{c_2}$ & Thresholds defining transitions between early, mid, and late experience stages \\
\hline
\end{tabular}
\label{tab:elo-notation}
\end{table}

In the Elo rating system, each player is assigned an initial rating $R_0$ upon registration. After every game, the ratings of both participants are updated using an Elo-based formula. Considering two players with current ratings \( rating_1 \) and \( rating_2 \), the updated ratings after a match are given by:

\[
\text{rating}_1' = \text{rating}_1 + K_1 \times (\text{Actual Score}_1 - \text{Expected Score}_1)
\]
and
\[
\text{rating}_2' = \text{rating}_2 + K_2 \times (\text{Actual Score}_2 - \text{Expected Score}_2)
\]

where \( K_1 \) and \( K_2 \) are sensitivity constants that depend on the number of games played by the respective players.

The expected score of player 1 against player 2 is modeled as:
\[
E_1 = \frac{1}{1 + 10^{\frac{\text{rating}_2 - \text{rating}_1}{D}}}
\]
and similarly, for player 2:
\[
E_2 = \frac{1}{1 + 10^{\frac{\text{rating}_1 - \text{rating}_2}{D}}}.
\]

The actual score of each player takes a value of 1, 0, or 0.5, corresponding to a win, loss, or draw, respectively. The sum of both players’ actual scores equals 1. D is a predetermined scaling factor.

The Elo model assumes that the difference in ratings provides a logistic estimate of expected performance. The scaling factor D is commonly chosen to be 400, such that a 400-point difference represents roughly 10:1 odds of victory for the stronger player, while a 200-point difference corresponds to approximately 3:1 odds.

To allow faster convergence of ratings during early games and stability in later stages, the \( K \)value is modeled as a monotonically non-increasing function of the number of games played by the $i^{th}$ player(\( n_i \)):

\begin{equation}
K_i =
\begin{cases}
K_a, & n_i \leq n_{c_1} \\
K_b, & n_{c_1} < n_i \leq n_{c_2} \\
K_c, & n_i > n_{c_2}
\end{cases}
\quad \text{for } i = 1, 2,\dots \text{with } K_a \geq K_b \geq K_c > 0
\label{k-values}
\end{equation}\\

This stepwise decay ensures that initial rating updates are more responsive, allowing a player’s rating to quickly approach their true skill level, while long-term stability is maintained as players play more games and more data becomes available. This method is applied by FIDE, and USCF (United States Chess Federation) using different values of $K$ depending on the skill-level (a combination of game count, current player rating, age) to create a more accurate and stable representation of a player's strength over time.

\section{Our Contribution: Parameter Tuning via Predictive Modeling}
\label{our_contribution}
In this section, we propose our data-driven tuning approach for determining optimal values $K_a, K_b, K_c$ and game-cutoff thresholds $n_{c_1}, n_{c_2}$, as denoted in \eqref{k-values} . We note that pre-match player ratings, or equivalently, rating difference (in a two-player game), are natural predictors of game outcome, as shown by Hvattum and Arntzen~\cite{football_pred}, where Elo-based ratings were assigned to teams and used as covariates to successfully predict outcomes of football matches, demonstrating the strong predictive power of rating systems. Based on this principle, player ratings are computed under multiple parameter configurations, and their predictive performance is evaluated using a classification model that estimates match outcomes from the initial rating difference between the two players. The parameter configuration maximizing predictive accuracy is selected by our method.  

The constants \( K_a, K_b, K_c, n_{c_1},  n_{c_2} \) were calibrated by computing player ratings under different parameter sets and evaluating their predictive power using an appropriate classification model.
In this study, logistic regression was chosen as the classification model due to its simplicity and ease of interpretation; however, other models such as random forests or neural networks can be applied similarly for the tuning task. 

Our approach introduces a robust technique for determining optimal parameter values in Elo’s foundational rating system~\cite{arpad_elo}, without any assumptions on the coefficients of player rating updates.  
Further, our method is independent of the specific rating formulation, whether Elo, Glicko, or any other system, as only the parameter space changes, while the optimization methodology remains similar.

\subsection{Choice of Parameters} 

In the traditional Elo system, three components influence rating updates: the expected score function $E$, the sensitivity constant $K$, and the scaling factor, typically $D = 400$. In our analysis, we do not modify $E$ since it is based on a distributional assumption on ratings, and the class of possible alternatives (based on suitable distributional assumptions) is too large for our method to be computationally feasible. While $D$ determines how differences in ratings translate to expected win probabilities, it can be noted that it is simply a scaling factor. Hence, changes in $D$ would only affect the dispersion of player ratings and not ranking or rating-based matchmaking. Hence, we consider a particular form of $E$ and constant $D$ in our implementation for ease of computation, though our method can be easily extended to considering these components as tunable parameters with feasible classes of alternatives.

In contrast, the $K$-factor directly affects rating responsiveness and stability by determining how much ratings change after each match. Further, since players have varying rates of learning a game, player ratings approach their true skill level at different rates depending on how many games they have played. So, we consider the K-factor and game thresholds for the K-factor as the parameters to be tuned using our method. 


\section{Empirical Analysis}
\label{empirical_results}
Our proposed empirical parameterization method was implemented on one synthetic and one real gameplay dataset to illustrate its practical usage. 
For this purpose, we used two main datasets, each containing game level data of the 3-dice, 2-player Ludo game. The rules of the game are available on the official Wowzy platform~\cite{wowzy_rules}. The real player data has been acquired from Games24x7.
\begin{itemize}
\item  The simulated gameplay dataset consists of a total of 184,000 games played between 7 different bots (or strategy profiles). The bot descriptions are given in section \ref{strategies}.
 \item The real player data consists of a total of 320,978 players who have played a total of 4,640,765 games among themselves across a time period of two and a half months in 2024. 
\end{itemize}

\subsection{Alternative Parameter Configurations}

In this subsection, we illustrate the parameter tuning approach employed in the empirical analysis, considering multiple configurations of score difference multipliers and cutoff thresholds. The K-factor for each game cutoff is denoted as $(K_a, K_b, K_c)$ as per the notations in \eqref{k-values}. The following choices of score difference multipliers were considered for our analysis. 

\begin{enumerate}
    \item (60, 30, 16): Baseline configuration used by major online chess systems.
    \item (30, 30, 30): Constant \( K \) used as a control setup for comparison.
    \item (30, 16, 8): Half of the base configuration, favoring higher stability.
    \item (100, 50, 25): Scaled-up version, providing higher responsiveness.
\end{enumerate}


A number of game cutoff thresholds were considered in the study, including both fixed and quantile-based thresholds: \((5, 10)\), \(([q_{10}] + 1, [q_{25}] + 1)\), and \(([q_{25}] + 1, [q_{50}] + 1)\), where \( q_k \) denotes the \(k^{th}\) percentile of the total game count distribution. 

Clearly, other parameter choices can also be considered as candidates for the optimal configuration. These specific values are primarily for demonstration of the parameter tuning method over a reasonably heterogeneous class of parameters.

The game cutoffs are expressed using a pair $(n_{c_1}, n_{c_2})$, as per the notations in \eqref{k-values}.

\vspace{0.5em}
\noindent

\subsubsection{Analysis on Real gameplay Data}

\begin{table}[H]
\centering
\begin{tabular}{ |c|c|c| }
\hline
 \textbf{K} & \textbf{Game Cutoffs} & \textbf{F1-Score} \\ 
\hline
 \rowcolor{yellow!50} \textcolor{blue}{60}, \textcolor{red}{30}, 16 & \textcolor{blue}{5}, \textcolor{red}{10} & 0.554 \\
 \textcolor{blue}{30}, \textcolor{red}{30}, 30 & \textcolor{blue}{5}, \textcolor{red}{10} & 0.548 \\
 \rowcolor{yellow!50} \textcolor{blue}{30}, \textcolor{red}{16}, 8 & \textcolor{blue}{5}, \textcolor{red}{10} & 0.554 \\ 
 \textcolor{blue}{100}, \textcolor{red}{50}, 25 & \textcolor{blue}{5}, \textcolor{red}{10} & 0.548 \\
\hline
 \rowcolor{yellow!50} \textcolor{blue}{60}, \textcolor{red}{30}, 16 & \textcolor{blue}{$[q_{10}] + 1$}, \textcolor{red}{$[q_{25}] + 1$} & 0.553 \\
 \textcolor{blue}{30}, \textcolor{red}{30}, 30 & \textcolor{blue}{$[q_{10}] + 1$}, \textcolor{red}{$[q_{25}] + 1$} & 0.548 \\
 \textcolor{blue}{30}, \textcolor{red}{16}, 8 & \textcolor{blue}{$[q_{10}] + 1$}, \textcolor{red}{$[q_{25}] + 1$} & 0.549 \\ 
 \textcolor{blue}{100}, \textcolor{red}{50}, 25 & \textcolor{blue}{$[q_{10}] + 1$}, \textcolor{red}{$[q_{25}] + 1$} & 0.550 \\
\hline
 \rowcolor{yellow!50} \textcolor{blue}{60}, \textcolor{red}{30}, 16 & \textcolor{blue}{$[q_{25}] + 1$}, \textcolor{red}{$[q_{50}] + 1$} & 0.554 \\
 \textcolor{blue}{30}, \textcolor{red}{30}, 30 & \textcolor{blue}{$[q_{25}] + 1$}, \textcolor{red}{$[q_{50}] + 1$} & 0.548 \\
 \textcolor{blue}{30}, \textcolor{red}{16}, 8 & \textcolor{blue}{$[q_{25}] + 1$}, \textcolor{red}{$[q_{50}] + 1$} & 0.551 \\ 
 \textcolor{blue}{100}, \textcolor{red}{50}, 25 & \textcolor{blue}{$[q_{25}] + 1$}, \textcolor{red}{$[q_{50}] + 1$} & 0.550 \\
\hline
\end{tabular}
\caption{F1-Score of Logistic Regression model for different values of constants $K_i$ and $n_i$}
\end{table}

Clearly, four of the chosen parameter configurations led to similarly high F1 scores. We proceed by choosing one of the configurations randomly, which leads to the following K - function \label{tuning-results}:
\[
K_i =
\begin{cases}
60, & n_i \leq 5 \\
30, & 5 < n_i \leq 10 \\
16, & n_i > 10
\end{cases}
\quad \text{for } i = 1, 2, ...
\]

After evaluation, we found that the logistic regression model exhibited modest predictive power for real gameplay. This can be attributed to factors such as:
\begin{itemize}
    \item The dataset represents early stages of the game’s release, when skill levels had not yet stabilized.
    \item The absence of prior experience data, which means that some “new” players may have been skilled from other platforms, introducing noise into the rating system.
\end{itemize}

To further corroborate this issue, we tried to find at what rating difference between the 2 players the model can start predicting well the winner of the game. 
\begin{table}[H]
\centering
\begin{tabular}{|l|l|l|l|l|}
\hline
\# & Rating Difference Range & Number of Games & Accuracy & F1-Score \\
\hline
0 & [0.0, 30.0] & 934529 & 0.52 & 0.36 \\
\hline
1 & (30.0, 60.0] & 824456 & 0.55 & 0.53 \\
\hline
2 & (60.0, 100.0] & 905680 & 0.59 & 0.57 \\
\hline
3 & (100.0, 200.0] & 1244662 & 0.66 & 0.63 \\
\hline
4 & (200.0, 500.0] & 329425 & 0.75 & 0.69 \\
\hline
5 & (500.0, 10000.0] & 5 & 0.8 & 0.67 \\
\hline
\end{tabular}
\caption{Accuracy and F1-Score across rating differences}
\end{table}

The results show that when the rating difference between two players is small (0–30), the model’s accuracy is close to random guessing (about 52\%), indicating the players are similarly skilled and outcomes are highly uncertain. As the rating difference increases, both accuracy and F1-score consistently improve. This suggests that once a noticeable skill gap emerges (beyond $\approx 60$ Elo), the higher-rated player is increasingly likely to win, and the model can predict outcomes with greater confidence. The predictive accuracy improves with increasing rating differences, reinforcing that rating disparities are a strong indicator of skill differences once the system stabilizes.
\\
The fitted logistic regression model yielded the following output:
\begin{table}[H]
\centering
\begin{tabular}{ |c|c|c|c|c| }
\hline
\textbf{Variable} & \textbf{Coefficient} & \textbf{Std. error} & \textbf{t-statistic} & \textbf{p-value}\\ 
\hline
Intercept & -0.0623  & 0.001 & -62.024 &  0.000  \\
\hline
 Rating Difference & 0.0046  & 9.49e-06 & 489.545 &  0.000  \\
\hline
\end{tabular}
\end{table}

The above table shows that rating difference is a statistically significant and positive predictor of winning probability.

This data-driven function effectively replaces the theoretical expected score \( E_1 \) in the Elo model, allowing empirical recalibration based on real or simulated match data.  
While the original Elo formulation uses a fixed scaling constant (400), this logistic regression implicitly learns an equivalent scaling from observed outcomes.  

Hence, changing the 400 constant in the Elo formula would modify the scaling of expected scores but not the underlying logistic relationship captured by the regression model.

\subsubsection{Analysis on Simulated gameplay data}
This section presents the Elo rating analysis on the simulated gameplay data, where the parameter configuration obtained using the parameter tuning exercise in \ref{tuning-results} was implemented. 

\begin{center}
\begin{table}[H]
\begin{tabular}{ |c|c|c|c|c| }
\hline
 \textbf{K} & \textbf{Game Cutoffs} & \textbf{F1-Score} & $\beta_0$ (Intercept) & $\beta_1$ (Rating difference coefficient) \\ 
\hline
 \textcolor{blue}{60}, \textcolor{red}{30}, 16 & \textcolor{blue}{5}, \textcolor{red}{10} & 0.87 & -0.132 & 0.0056 \\
\hline
\end{tabular}
\end{table}
\end{center}

The F1 scores obtained on the simulated dataset are consistently higher than those observed on real gameplay data, due to the controlled nature of simulated matches. In simulations, strategy profiles make consistent decisions, leading to better skill distinctions. As a result, the relationship between skill difference and match outcomes is stronger and easier for the Elo-based model to learn and predict accurately.

In contrast, real player data introduces several sources of noise, such as variations in human decision-making, inconsistent strategies over time, and psychological factors. Furthermore, players occasionally outperform or under-perform relative to their true skill level, making win/loss outcomes less deterministic. This increases both false positives and false negatives in prediction, thereby lowering the F1 score.

Thus, the higher F1-score on simulated data reflects a more stable and structured competitive environment, whereas the drop in F1-score for real data highlights the inherent uncertainty and variability in human gameplay.

The logistic regression model estimated the probability of Player 1 winning a match as a function of the rating difference between the two players. 
\\
The fitted model provided the following results:
\begin{table}[H]
\centering
\begin{tabular}{ |c|c|c|c|c| }
\hline
\textbf{Variable} & \textbf{Coefficient} & \textbf{Std. error} & \textbf{t-statistic} & \textbf{p-value}\\ 
\hline
Intercept & -0.1298  & 0.008 & -16.373 &  0.000  \\
\hline
Rating Difference & 0.0056  & 2.56e-05 & 219.460 &  0.000  \\
\hline
\end{tabular}
\end{table}


The results in this section validate the expected monotonic relationship between rating difference and win probability, consistent with the foundations of the Elo system.

Future work may extend this by modelling it \( K \) as a function of the player’s current rating rather than the number of games played, similar to the dynamic approach adopted by FIDE in recent years.

\section{Illustration of Empirical Results}
\label{illustration_of_results}

This section presents the details of the implemented rating system based on the empirical parameterisation technique.

\subsection{Analysis on Real gameplay data}



\begin{table}[H]
\centering
\begin{tabular}{|c|c|c|c|c|c|c|c|c|}
\hline
\textbf{Statistic} & Min & 10\% & 25\% & 50\% & Mean & 75\% & 90\% & Max \\ 
\hline
\textbf{Value} & 776 & 977 & 1028 & 1094 & 1096 & 1162 & 1216 & 1452 \\
\hline
\end{tabular}
\caption{Summary statistics of dynamic ludo ratings}
\end{table}

The ratings span a wide range from 776 to 1452, with a near-symmetric distribution centered close to the initial rating of 1000. 

\begin{figure}[H] 
    \centering
    \includegraphics[height = 6 cm, width = 10 cm]{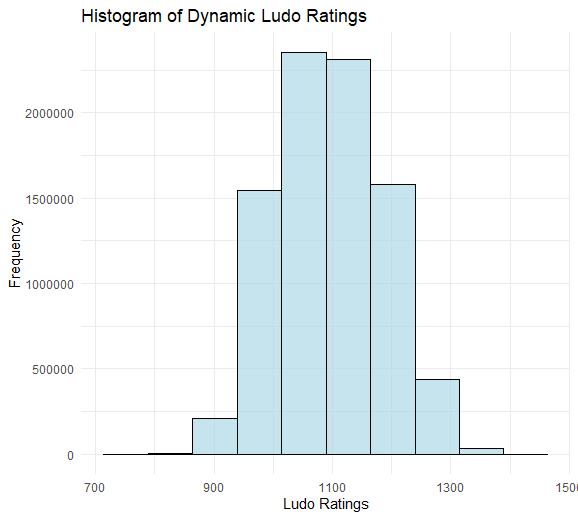} 
    \caption{Histogram of Ludo Dynamic Ratings}
    \label{fig:ludo_ratings_hist} 
\end{figure}

The dynamic Ludo ratings follow a bell-shaped distribution, in line with the assumptions of the Elo rating system. Most ratings are concentrated near the central range (1000–1150), suggesting that the Elo updating mechanism maintains equilibrium across the player population, with few extreme outliers.



\begin{figure}[H] 
    \centering
    \includegraphics[width = 1.1\textwidth]{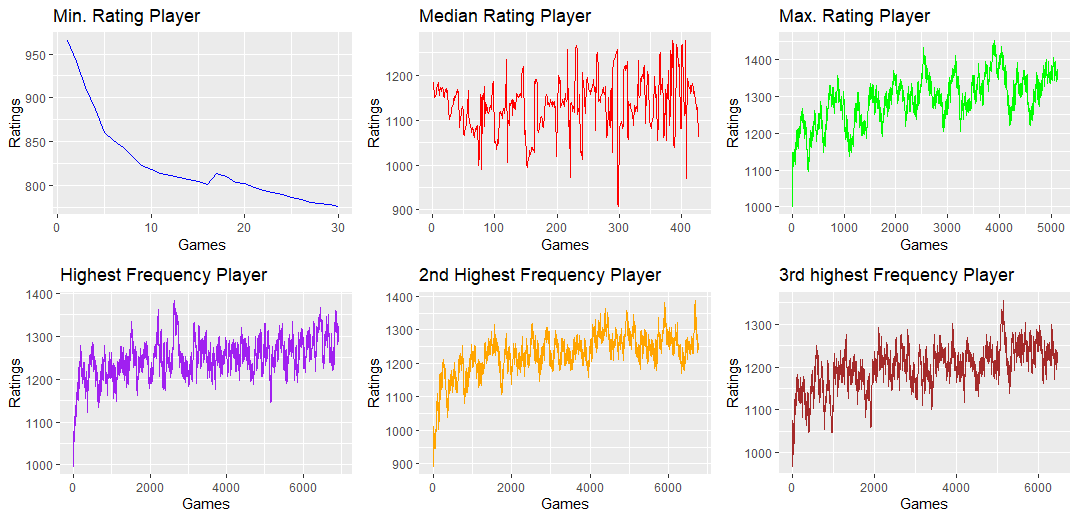} 
    \caption{Dynamic Ludo Rating Curves of Certain Players} 
    \label{fig:ludo_dynamic_ratings_splot} 
\end{figure}

The rating trajectories of the top-rated and high-frequency players display a consistent upward trend, indicating progressive skill improvement with more gameplay. In contrast, the median-rated player exhibits significant fluctuations around the baseline rating (1000), stabilising slightly above it after repeated play. The lowest-rated player, with far fewer games, shows a clear downward trajectory, suggesting that limited gameplay experience may contribute to poorer performance and slower rating recovery.

\subsection{Analysis of Simulated gameplay data}

The simulated Ludo dataset consists of 184,000 games played between 7 bots of varying strength levels, as defined in \ref{strategies}. Match-ups were assigned randomly to avoid pairing bias.

The bots are divided into two categories – Set I (Aggressive, Responsible, Naive), which follow pre-defined decision-making algorithms, and Set C (Full Information, Limited Information, Random, and Defeat Seeking), which makes learning-based decisions (Monte Carlo Tree Search)~\cite{informedMCTS2016, mcts_paper}. The complete details of the strategies are present in section \ref{strategies} below. For our purpose, it suffices to mention that Full Information, Aggressive and Responsible Pair are the more sophisticated algorithms with ``better" decision-making rules compared to the other bots.

The strongest strategies across both sets—Full Information, Aggressive, and Responsible Pair—exhibit nearly identical performance levels, and clearly outperform the weaker strategies. This is clearly reflected in both the final rating values as well as the rating progression curves.

\begin{table}[H]
\centering
\begin{tabular}{|c|c|}
\hline
\textbf{Bot} & \textbf{Final Rating}  \\ 
\hline
\rowcolor{yellow} Aggressive & 1391  \\ 
\hline
Naive  & 1007 \\ 
\hline
\rowcolor{yellow} Responsible Pair  & 1393\\ 
\hline
\rowcolor{yellow} Full Information & 1385 \\ 
\hline
Limited Information & 993 \\ 
\hline
Defeat Seeking & 60 \\ 
\hline
Random & 775 \\ 
\hline
\end{tabular}
\caption{Final Ratings of Simulated Gameplay Data}
\label{bot_final_rating}
\end{table}

\begin{figure}[H] 
    \centering
    \includegraphics[width=0.75\textwidth]{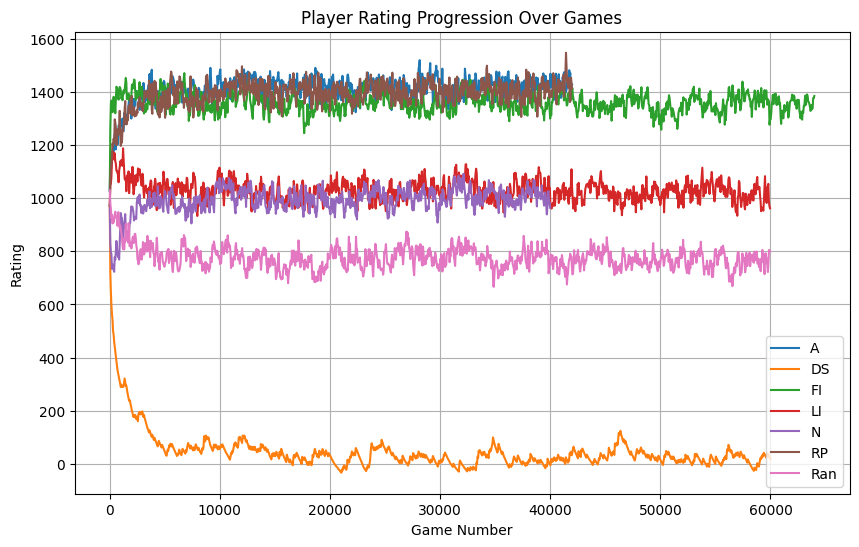} 
    \caption{Rating progression of each strategy implemented bot}
    \label{fig:ludo_ratings_simualted} 
\end{figure}

Bots employing strategically advanced decision-making (Aggressive, Responsible Pair, and Full Information) consistently converge to higher Elo ratings ($\approx 1380–1400$), demonstrating superior long-term performance. In contrast, weaker or irrational strategies such as Naive, Random, and especially Defeat Seeking, stabilize at substantially lower ratings, confirming their competitive disadvantage.

The dynamic rating curves in Figure~\ref{fig:ludo_ratings_simualted} show three important characteristics:

\begin{itemize}
    \item Rating stabilization: After approximately 5000 games, ratings begin to stabilize, indicating that the system has accumulated sufficient evidence to reflect true relative skill levels.
    \item Strategy separability: Stronger strategies remain clearly separated from weaker ones over time, demonstrating rating sensitivity to decision-making quality.
    \item Convergence of weak strategies: The Defeat Seeking bot shows rapid decay and stabilizes close to zero, validating that the rating system strongly penalizes consistently poor decision-making.
\end{itemize}


The dynamic $K$-factor further accelerates this convergence by assigning higher update sensitivity during the early stages of gameplay when uncertainty is high and reducing it as more data becomes available. This property makes Elo especially suitable for new platforms where player histories are initially sparse.

These results demonstrate that the rating model successfully distinguishes strategies by skill and reaches stable estimates when exposed to sufficient gameplay data.

\section{Conclusion}
\label{conclusion}
The data-driven parameterization technique of the Elo rating system proposed in this paper puts forward a robust method of measuring player performance in 2-player games. Unlike traditional rating systems, which rely on fixed or heuristic parameter choices, this method optimizes parameters empirically based on predictive accuracy, allowing the rating system to better reflect true player performance. This is further highlighted from the implementation of the rating system on real and synthetic gameplay datasets, which shows accurate modelling of player performance based on the empirically parameterized rating system. Further, the model is easily extensible to other rating systems and parameter configurations with suitable modifications.

This work establishes a foundation for robust modelling of player skill utilizing  gameplay data. There are many promising possibilities for future research in this domain, including parameter tuning based on combinations of model outputs for different classification models and comparison of parameter configurations across rating systems. Additionally, leveraging existing game ecosystems to define data-driven baseline configurations may lead to the development of new rating systems, considerably impacting the problem of player skill quantification in sports analytics.

\section*{Appendix: Specification of Players' Strategies for Empirical Evaluation}
\label{strategies}

\newenvironment{strategy}[1]
  {\par\noindent\textbf{#1:} }
  {\par\medskip}

\newenvironment{SetI}
  {\subsection*{Set I Strategies -}\par}
  {\par\medskip}
  
\newenvironment{SetC}
  {\subsection*{Set C Strategies -}\par}
  {\par\medskip}
  
For the empirical evaluation, we create two sets of different alternatives (a total of seven) that represent different decision capabilities while playing the game. These sets are: (i) \textbf{Set I} --- \textit{intuitive rule based} strategies, similar to those mentioned in the previous section; and (ii) \textbf{set C} --- \textit{algorithmic} strategies, using either random or sophisticated self-learning based game-play using Monte-Carlo-Tree-Search (MCTS)~\cite{chaslot2008mcts}. In this section, we describe these strategies in detail.

\begin{SetI}

We use three alternatives, where one is similar to (PP) (called ``Aggressive"), a second similar to (S) (called ``Responsible Pair") and a relatively unsophisticated algorithm (called ``Naive"). We detail below.

\begin{strategy}{The ``Naive" Player (N)}

The Naive player follows a very simple algorithm for movement of tokens.  As long as the first token is active, it is moved using the first available dice roll; unless the move is illegal. \\ If the first token is inactive or its movement is illegal, the next active token is moved similarly. \\
Further, the capture of a token does not affect its movement priority.
\end{strategy}
\begin{strategy}{The ``Aggressive" Player (A)}

Designed similarly to the PP strategy, the Aggressive player aims to promote tokens one at a time, prioritizing captures.
The action performed by the Aggressive player in any turn (if possible), in decreasing order of priority is: (a) promoting a token, (b) capturing an opponent token, (c) moving a token to a safe square, (d) moving the first active token with the highest available roll.
\end{strategy}

\begin{strategy}{The ``Responsible Pair" Player (RP)}

The ``Responsible Pair" player prioritizes safe play, initially aiming to reach the opponent's starting square. It also permits situational aggressive movement.
The action performed by the Responsible Pair player in a turn, in decreasing order of priority is: (a) promoting a token, (b) capturing an opponent token, (c) \textit{situational aggressive movement - }moving the highest point token when the opponent's token is close to promotion, (d) moving a token to a safe square, (e) chasing an opponent token with either of its last two tokens, (f) moving the tokens alternately in pairs till they reach 27 with the highest available roll, (g) moving the tokens alternately in pairs after all of them have reached 27 with the highest available roll.
\end{strategy}
\end{SetI}
\vspace{-5mm}
\begin{SetC}
\begin{strategy}{The ``Full-Information" Player (FI)}

The FI player generates decision trees corresponding to each action at a game state. Any specific turn of the game permits a maximum of 12 possible actions (3 dice multiplied by the 4 movable tokens). Each branch (possible action) is then taken, and the game is simulated to its conclusion, akin to Monte Carlo Tree Search (MCTS)~\cite{mcts_paper} for 100 iterations to determine the win rate for each branch. The action with the highest win rate (based on the posterior simulation) is selected for movement.

\end{strategy}

\begin{strategy}{The ``Limited-Information" Player (LI) }

The LI player adopts a strategy similar to the FI player but limits its decision trees to the first available roll. Thus, it permits a maximum of 4 possible actions and chooses the action with the highest win rate (based on posterior simulation) for movement.

\end{strategy}

\begin{strategy}{The ``Defeat-Seeking" Player (DS) }

The DS player mirrors the approach of the Full-Information bot but inversely, opting for the dice and token combination that results in the lowest win rate in the posterior simulations.

\end{strategy}

\begin{strategy}{The ``Random" Player (R)}

The Random player chooses one of the four tokens and one of the available rolls independently with equal probability for token movement, provided the resultant move is legal. For example, given dice rolls [2, 5,\_] and randomly choosing dice 5, with token positions at [54, 2, 3, 3], the bot randomly selects a token from the movable tokens (in this case, tokens 2, 3, and 4).

\end{strategy}
\end{SetC}

\end{document}